\documentclass[conference]{IEEEtran} 

\usepackage{cite}
\usepackage{amsmath,amssymb,amsfonts}
\usepackage{algorithmic}
\usepackage{graphicx}
\usepackage{textcomp}
\usepackage{xcolor}
\usepackage{mathtools}
\def\BibTeX{{\rm B\kern-.05em{\sc i\kern-.025em b}\kern-.08em
    T\kern-.1667em\lower.7ex\hbox{E}\kern-.125emX}}

\usepackage{algorithm}
\usepackage{algorithmic}

\usepackage{color}
\usepackage{tablefootnote}

\usepackage{amsmath,amssymb,amsthm}
\usepackage{multirow}

\newcommand{\mbf}{\mathbf}

\usepackage{algorithm}
\usepackage{algorithmic}
\usepackage{longtable,tabularx}

\begin{document}

\title{Advances in Particle Flow Filters with Taylor Expansion Series}

\author{

\IEEEauthorblockN{Simone Servadio,}
\IEEEauthorblockA{\textit{Iowa State University}, Ames, IA, USA}
\thanks{Dr. Simone Servadio, Assistant Professor, Department of Aerospace Engineering, servadio@iastate.edu}

}

\markboth{Transactions on Aerospace and Electronic Systems}%
{Shell \MakeLowercase{\textit{et al.}}: Bare Demo of IEEEtran.cls for IEEE Journals}

\maketitle
\thispagestyle{plain}
\pagestyle{plain}

\begin{abstract}
Particle Flow Filters perform the measurement update by moving particles to a different location rather than modifying the particles' weight based on the likelihood. Their movement (flow) is dictated by a drift term, which continuously pushes the particle toward the posterior distribution, and a diffusion term, which guarantees the spread of particles. This work presents a novel derivation of these terms based on high-order polynomial expansions, where the common techniques based on linearization reduce to a simpler version of the new methodology. Thanks to differential algebra, the high-order particle flow is derived directly onto the polynomials representation of the distribution, embedded with differentiation and evaluation. The resulting technique proposes two new particle flow filters, whose difference relies on the selection of the expansion center for the Taylor polynomial evaluation. Numerical applications show the improvement gained by the inclusion of high-order terms, especially when comparing performance with the Gromov flow and the ``exact" flow. 
\end{abstract}


\IEEEpeerreviewmaketitle

%%%%%%%%%%%%%%%%%%%%%%%%%%%%%%%%%%%%%%%%%%%%%%%%%%%%%%%%%%%%%%%%%%%%%%%%%%%%%%%%%%%%%%%%%
%%%%%%%%%%%%%%%%%%%%%%%%%%%%%%%%%%%%%%%%%%%%%%%%%%%%%%%%%%%%%%%%%%%%%%%%%%%%%%%%%%%%%%%%%
%%%%%%%%%%%%%%%%%%%%%%%%%%%%%%%%%%%%%%%%%%%%%%%%%%%%%%%%%%%%%%%%%%%%%%%%%%%%%%%%%%%%%%%%%

\section{Introduction}
\IEEEPARstart{O}{ne} of the main challenge in estimation and measurement fusion is to accurately represent the posterior Probability Density Functions (PDF) of the state of a system, which is derived after obtaining an external observation from the sensors and performing the measurement update. 

The PDF representation with particles, regardless of its complexity, is one of the most accurate selection to approximate the true posterior. Traditionally, the most common choice lands on Sequential Monte Carlo methods, which are nonlinear estimators where PDFs are approximated as Probability Mass Function (PMF) using a set of
particles \cite{smith2024uncertainty}. The Bootstrap Particle Filter (BPF) \cite{gordon1993novel} assigns particle weights according to the measurement likelihood, which can lead to particle collapse or degeneracy when an unlucky observation is received. Sequential Importance Samplings (SIS) help draw particles near high-density probability regions to achieve an accurate estimate. The SIS is based on a smart selection of the importance distribution, such as the Unscented Particle Filter (UPF) \cite{van2000unscented} where the updated Unscented Kalman Filter (UKF) \cite{julier2004unscented} distribution is used to sample particles.  

On the contrary, Particle Flow Filtering (PFF) is based on the idea of keeping the weight of each particle constant, performing the measurement update by moving them to new regions. Thus, by definition, PFF is immune to particle collapse and degeneracy. Daum and Huang \cite{daum2007nonlinear} showed the log-homotopy that transforms the prior distribution into the posterior and that identifies the ``flow" of particles. Initially, they derived the \textit{exact} flow, a zero-diffusion solution for PFF \cite{daum2010exact}. Afterwards, the Gromov flow was introduced, which includes a diffusion term \cite{daum2016gromov}. Crouse and Lewis present a complete derivation of the various flows, analyzing the different solutions and integration techniques \cite{crouse2019consideration}. Recently, a new particle flow has been derived as a recursive update, where a set of stochastic differential equations describe the flow of the particles \cite{michaelson2024particle}. 

This paper reconstructs the flow from its original derivation before the introduction of any Gaussian or linearization assumption, likewise in the exact and Gromov flow. From this general representation, the flow is derived in the Differential Algebra (DA) framework to account for high-order term contributions rather than stopping at the mere linearization. The use of DA in estimation has been proven successful by representing PDF directly as polynomials \cite{servadio2022maximum} or via central moments \cite{servadio2022nonlinear}. In the paper, DA is used for the first time to enhance PFF, leveraging its computational benefits and higher accuracy, deriving the Differential Algebra Particle Flow Filter (DAPFF).

%%%%%%%%%%%%%%%%%%%%%%%%%%%%%%%%%%%%%%%%%%%%%%%%%%%%%%%%%%%%%%%%%%%%%%%%%%%%%%%%%%%%%%%%%
%%%%%%%%%%%%%%%%%%%%%%%%%%%%%%%%%%%%%%%%%%%%%%%%%%%%%%%%%%%%%%%%%%%%%%%%%%%%%%%%%%%%%%%%%
%%%%%%%%%%%%%%%%%%%%%%%%%%%%%%%%%%%%%%%%%%%%%%%%%%%%%%%%%%%%%%%%%%%%%%%%%%%%%%%%%%%%%%%%%
\section{Measurement Fusion via Bayesian Update} \label{sec1}
Given some ``a priori" knowledge of the state PDF, $p_{\mbf x}(\mbf x)$ the \textit{prior} distribution, and a measurement $\mbf y$ with known relationship to the state, the ``a posteriori" distribution, $p_{\mbf x}(\mbf x|\mbf y)$ the \textit{posterior} distribution, specifies the probability density of the state given the measurement $\mbf y$. The evaluation of the conditional distribution follows Bayes' rule, which is written as 
\begin{equation}
    p_{\mbf x}(\mbf x|\mbf y) = \dfrac{p_{\mbf x}(\mbf x) p_{\mbf y}(\mbf y|\mbf x)}{p_{\mbf y}(\mbf y)}
\end{equation}
where $p_{\mbf y}(\mbf y|\mbf x)$ is the conditional distribution of the measurement given the state, called \textit{likelihood}, and $p_{\mbf y}(\mbf y)$ is the normalizing constant, the marginal distribution of the measurement, such that the posterior integrates to unity in its domain. After taking the log function, the Bayes's formulation becomes
\begin{equation}
    \log p_{\mbf x}(\mbf x|\mbf y) = \log p_{\mbf x}(\mbf x) + p_{\mbf y}(\mbf y|\mbf x) - \log p_{\mbf y}(\mbf y)\label{log_Bayes}
\end{equation}
For traditional particle filters, such as the Bootstrap Particle Filter (BPF) \cite{arulampalam2002tutorial}, the PDF is approximated by an ensemble of particles whose weight changes according to their likelihood of producing the given measurement. That is, after propagation and prediction, the current weight of each particle $\omega^{(i)}_{prior}$ gets scaled based on the likelihood function in the update step of the filter
\begin{equation}
    \omega^{(i)}_{post} \propto p_{\mbf y}(\mbf y|\mbf x^{(i)}) \omega^{(i)}_{prior}
\end{equation}
where proportionality is used over equality as it is assumed that weights get normalized to sum to unity after multiplication. This approach is sometimes not very effective, as it can cause particle degeneracy, where a few samples get all the weight, making this representation of the PDF faulty. Even with resampling, the space of the posterior PDF cannot be accurately represented if only a small percentage of the ensemble has a significant portion of the overall weight. Therefore, a different approach has been implemented \cite{daum2007nonlinear} where particles move during the update step rather than changing their weight. The idea is to shift the particles according to the likelihood distribution to better represent the posterior PDF, moving as flow. Thus, particle flow filters are immune to particle degeneracy as their weight is constant.

%%%%%%%%%%%%%%%%%%%%%%%%%%%%%%%%%%%%%%%%%%%%%%%%%%%%%%%%%%%%%%%%%%%%%%%%%%%%%%%%%%%%%%%%%
%%%%%%%%%%%%%%%%%%%%%%%%%%%%%%%%%%%%%%%%%%%%%%%%%%%%%%%%%%%%%%%%%%%%%%%%%%%%%%%%%%%%%%%%%
%%%%%%%%%%%%%%%%%%%%%%%%%%%%%%%%%%%%%%%%%%%%%%%%%%%%%%%%%%%%%%%%%%%%%%%%%%%%%%%%%%%%%%%%%
\section{Particle Flow Filtering } 
The main idea of particle flow is to update the distribution iteratively. Starting from Eq. \eqref{log_Bayes}, particle flow defines a scalar homotopy parameter $0\leq \lambda \leq 1 $ and the relative log-homotopy in the form of:
\begin{equation}
    \log p_{\mbf x}(\mbf x|\mbf y, \lambda) = \log p_{\mbf x}(\mbf x) + \lambda \log p_{\mbf y}(\mbf y|\mbf x) - \log K(\lambda) \label{log_bayes}
\end{equation}
where the normalization constant $K$ has been parametrized by $\lambda$ to ensure that the posterior distribution is a valid PDF that integrates to unity. It is evident that $p_{\mbf x}(\mbf x|\mbf y, 0)$ reduces to the prior distribution while $p_{\mbf x}(\mbf x|\mbf y, 1)$ is the posterior PDF, as it represents the classic Baye's formulation.

The main idea of particle flow is to consider the parameter $\lambda$ as a pseudo-time in which the prior is transformed into the posterior. Therefore, the update step of the particle filter happens in this pseudo-time domain, following a stochastic differential equation that defines the flow of the particles:
\begin{equation}
    d \mbf x = \mbf f (\mbf x,\lambda) d\lambda + \mbf B(\mbf x,\lambda)d\mbf w_{\lambda}
\end{equation}
where $\mbf f (\mbf x,\lambda)$ is the drift function and $\mbf B (\mbf x,\lambda)$ is the diffusion matrix function. These functions depend both on the state $\mbf x$ and the pseudo-time $\lambda$, while $d\mbf w_{\lambda}$ is a Weiner process that accounts for the stochastic nature of the differential equation, making the path of the particle during the update random. If $\mbf B (\mbf x,\lambda)$ is always null, then the flow becomes deterministic, and the particle follows the differential equation
\begin{equation}
    \dfrac{d \mbf x}{d \lambda} = \mbf f (\mbf x,\lambda)
\end{equation}

These stochastic differential equations can be solved similarly to uncertainty propagation, using the Fokker-Plank equation. 
\begin{align}
    \dfrac{\partial p_{\mbf x}(\mbf x, \lambda)}{\partial \lambda} =&  - \nabla_{\mbf x} \cdot \left(p_{\mbf x}(\mbf x, \lambda)\mbf f (\mbf x,\lambda) \right)+\nonumber \\
    +& \dfrac{1}{2}\nabla^T_{\mbf x} \left(p_{\mbf x}(\mbf x, \lambda)\mbf Q (\mbf x,\lambda) \right)\nabla_{\mbf x}
\end{align}
where $\nabla_{\mbf x} \cdot$ is the divergence operator, $\nabla_{\mbf x}$ represents the gradient operator, and where $\mbf Q(\mbf x,\lambda) =  \mbf B(\mbf x,\lambda) \mbf B^T(\mbf x,\lambda)$

The derivation of the flow, and the solution of the drift and diffusion terms, is cumbersome and reported in \cite{crouse2019consideration}, as well as in \cite{daum2016gromov}. Working with logarithmic identities leads to the solution of the drift term as
\begin{equation}
    \mbf f(\mbf x, \lambda) = - \left( \nabla_{\mbf x} \nabla_{\mbf x} ^T \log P (\mbf x, \lambda)^{} \right)^{-1} \nabla_{\mbf x} \log L(\mbf x, \lambda) \label{drift_gen}
\end{equation}
and the solution of the diffusion term as
\begin{equation}
    \mbf Q(\mbf x, \lambda) = \left( \nabla_{\mbf x} \nabla_{\mbf x} ^T \log P (\mbf x, \lambda) \right)^{-1} \nabla_{\mbf x} \mbf f^T(\mbf x, \lambda) \label{diff_gen}
\end{equation}
where $L(\mbf x) = p_{\mbf y}(\mbf y|\mbf x)$ is the likelihood function and $P (\mbf x) = p_{\mbf x}(\mbf x|\mbf y)$ is the posterior distribution. 

These equations, Eq. \eqref{drift_gen} and Eq.\eqref{diff_gen}, are the general solution to the flow of particles, where the prior PDF is a generic distribution and the measurement equation is nonlinear and affected by nongaussian noise, according to 
\begin{equation}
    \mbf y = \mbf h(\mbf x) + \mbf v
\end{equation}
where $\mbf h(\mbf x)$ is the nonlinear measurement equation affected by the nongaussian noise $ \mbf v$, with known covariance matrix $\mbf R$.

%%%%%%%%%%%%%%%%%%%%%%%%%%%%%%%%%%%%%%%%%%%%%%%%%%%%%%%%%%%%%%%%%%%%%%%%%%%%%%%%%%%%%%%%%
\subsection{The Gromov Flow}
The Gromov flow is a flow with non-zero diffusion, and assumes a prior Gaussian distribution with linear and Gaussian measurements. Under these assumptions, the equations from the general case can be derived analytically. In practice, for nonlinear nongaussian measurements, the gradient of the measurement function evaluated at each particle location is employed, approximating the noise as Gaussian with the given covariance matrix $\mbf R$. 

Therefore, after defining the measurement Jacobian as 
\begin{equation}
    \mbf H = \nabla_{\mbf x} \mbf h(\mbf x) 
\end{equation}
Eq. \eqref{drift_gen} and Eq. \eqref{diff_gen} are solved and lead to the classic Gromov equations. To simplify nomenclature, let us define matrix $\mbf S$ as 
\begin{equation}
    \mbf S = (\mbf P^{-1}+\lambda\mbf H^T \mbf R^{-1} \mbf H)^{-1} 
\end{equation}
which is similar to the information matrix, where $\mbf P$ indicates the current state covariance matrix. Thus, the drift and diffusion terms can be expressed as
\begin{align}
    \mbf f(\mbf x, \lambda) &= - \mbf S \mbf H^T\mbf R^{-1} (\mbf h(\mbf x)-\mbf y) \label{drift_gro}\\
    \mbf Q (\lambda) &= \mbf S \mbf H^T\mbf R^{-1} \mbf H \mbf S\label{diff_gro}
\end{align}
where it can be noted that, as expected, the diffusion term is the same regardless of the state, meaning that each particle is assigned the same random spread, while the drift term depends on where the particle resides, since it must move accordingly. These two equations are solved for each particle $\mbf x^{(i)} $, integrating the pseudo-time $\lambda $ from 0 to 1. 

Given $\mbf Q (\lambda)$, it is necessary to take the matrix square root to obtain $\mbf B(\lambda)$ such that $\mbf Q(\lambda) =  \mbf B(\lambda) \mbf B^T(\lambda)$. This step requires attention as $\mbf Q (\lambda)$ can easily become singular, meaning that a Cholesky decomposition would fail in providing a lower triangular matrix and other techniques, such as LDL decomposition, must be used. 

%%%%%%%%%%%%%%%%%%%%%%%%%%%%%%%%%%%%%%%%%%%%%%%%%%%%%%%%%%%%%%%%%%%%%%%%%%%%%%%%%%%%%%%%%
\subsection{The Exact Flow}
The exact flow is a zero diffusion flow that uses an explicit solution of the update differential equation while neglecting the normalization term of the flow and approximating the prior and measurement distributions as Gaussian. The idea is to use the same weight particles to represent the shape of the posterior as accurately as possible, given the simplifying assumptions of Gaussianity. Once again, in the case of a nonlinear measurement function, localized linearization at each particle's location is implemented. The drift term is expressed as the summation of two components for each particle. Mallik \cite{mallick2015critical} has a step-by-step derivation of the flow, which can be summarized as:
\begin{align}
    \mbf f (\mbf x^{(i)}, \lambda) &= \mbf A(\lambda) \mbf x^{(i)} + \mbf b (\lambda) \\
    \mbf r &= \mbf y - \mbf h(\mbf x^{(i)}) + \mbf H \mbf x^{(i)} \\
    \mbf A(\lambda) &= \dfrac{1}{2} \mbf P \mbf H^T \left( \lambda \mbf H \mbf P \mbf H^T + \mbf R\right)^{-1}\mbf H \\
    \mbf b(\lambda) &= (\mbf I + 2\lambda\mbf A) \big(\mbf A\mbf {\hat x}^- + (\mbf I + \lambda \mbf A)\mbf P \mbf H^T \mbf R^{-1} \mbf r \big)
\end{align}
where, $\mbf {\hat x}^-$ is the prior mean and $\mbf P$ is the state covariance matrix. These equations repeat themselves recursively for each particle and for each value of $\lambda$ until it reaches unity. The particles' and covariance's dependency on $\lambda$ has been omitted to provide cleaner formulations. The need for the state covariance matrix requires this particle flow filter to work alongside an EKF or UKF, otherwise it can evaluate the covariance directly from the particles at the end of each measurement update.

%%%%%%%%%%%%%%%%%%%%%%%%%%%%%%%%%%%%%%%%%%%%%%%%%%%%%%%%%%%%%%%%%%%%%%%%%%%%%%%%%%%%%%%%%
%%%%%%%%%%%%%%%%%%%%%%%%%%%%%%%%%%%%%%%%%%%%%%%%%%%%%%%%%%%%%%%%%%%%%%%%%%%%%%%%%%%%%%%%%
%%%%%%%%%%%%%%%%%%%%%%%%%%%%%%%%%%%%%%%%%%%%%%%%%%%%%%%%%%%%%%%%%%%%%%%%%%%%%%%%%%%%%%%%%
\section{The Differential Algebra Approach}
The main idea of the advantages of including DA in the particle flow mathematics comes from relaxing the linearization constraint. The Gromov's and exact flow's drift and diffusion terms derive the equations considering a linear approximation of the measurement function around each particle. The use of differential algebra aims at expanding such approximation to larger, higher orders, using an algebra of Taylor polynomials. That is, the derivations from Eq. \eqref{drift_gen} and Eq.\eqref{diff_gen} can be computed with a higher accuracy level when embedded in the DA framework. 

Therefore, in this novel derivation, every variable, function, and distribution is represented in its Taylor polynomial expansion series up to an arbitrary order. Thus, the prior distribution, likelihood, and posterior are represented in their polynomial form as 
\begin{align}
    \mathcal{T}(\mbf x) &= \log p(\mbf x) \\
    \mathcal{L}(\mbf x) &= \log p(\mbf y|\mbf x) \\
    \mathcal{P}(\mbf x, \lambda) &= \log p(\mbf x|\mbf y, \lambda)
\end{align}
Using this definition, the drift term in the DA framework can be calculated directly onto the polynomial as is Eq. \eqref{drift_gen}
\begin{align}
    \boldsymbol{\mathcal{F}}(\mbf x, \lambda) &= - \left( \nabla_{\mbf x} \nabla_{\mbf x} ^T \mathcal{P} (\mbf x,\lambda) \right)^{-1} \nabla_{\mbf x} \mathcal{L}(\mbf x) \\
    &= \text{Hess}(\mathcal{P} (\mbf x,\lambda))^{-1} \nabla_{\mbf x} \mathcal{L}(\mbf x) \label{drift_DA} 
\end{align}
This is a vector like $\mbf f(\mbf x,\lambda)$ where each component is a polynomial, where $\text{Hess}()$ represents the Hessian of the function. In a similar manner, the diffusion term is evaluated as is Eq. \eqref{diff_gen}
\begin{align}
    \boldsymbol{\mathcal{Q}}(\mbf x, \lambda) &= \left( \nabla_{\mbf x} \nabla_{\mbf x} ^T \mathcal{P} (\mbf x,\lambda) \right)^{-1} \nabla_{\mbf x} \mathcal{L} (\mbf x) \\
    &= \text{Hess}(\mathcal{P} (\mbf x,\lambda))^{-1} \nabla_{\mbf x} \boldsymbol{\mathcal{F}}^T(\mbf x, \lambda)  \label{diff_DA}
\end{align}
This is a matrix like $\mbf Q(\lambda)$, where each entry is a polynomial. 

The feasibility of the approach is based on the differentiability of the measurement function, as multiple terms must be computed to increase the accuracy of the measurement update depending on the orders of the expansion series. The other selection that must be made is the expansion center, i.e., the location at which the derivatives are calculated to provide the coefficients of the terms in the polynomial. Depending on the selection, two separate particle flow update techniques have been developed in the DA framework. The first one, the Differential Algebra Particle Flow Filter version 1 (DAPFFv1-$c$), selects the mean of the prior distribution as the center of the expansion; the second one, the Differential Algebra Particle Flow Filter version 2 (DAPFFv2-$c$), expands a polynomial map centered at each particle's location. The integer number $c$ after the name of the filter specifies the order of the expansion series, which is selected arbitrarily.

%%%%%%%%%%%%%%%%%%%%%%%%%%%%%%%%%%%%%%%%%%%%%%%%%%%%%%%%%%%%%%%%%%%%%%%%%%%%%%%%%%%%%%%%%
\subsection{The Prior Mean as the Expansion Center - DAPFFv1-$c$}
In this first version of the DA particle flow update, the mean on the prior distribution, $\hat{\mbf x}^-$ has been chosen as the expansion center of the Taylor polynomial map. Thus, the state polynomial is initialized as a function of deviations
\begin{equation}
    \mbf x(\delta \mbf x) =  \hat{\mbf x}^- + \delta \mbf x
\end{equation}
where $\delta \mbf x$ is the deviation of each particle from the mean. This representation means that each particle $\mbf x^{(i)}$ has associated its own deviation $\delta \mbf x^{(i)}$ such that 
\begin{equation}
    \delta \mbf x^{(i)} = \mbf x^{(i)} - \hat{\mbf x}^-
\end{equation}
The prior distribution covariance, $\mbf P^-$ is either given or evaluated directly by the particles at their current location. The Gromov's flow assumes a Gaussian prior and derives the drift and diffusion equations with the knowledge of the prior mean and covariance. The DA approach expands on this assumption by working directly onto the prior distribution, taking advantage of the logarithm applied in Eq. \eqref{log_bayes}. Therefore, using the polynomial representation of the state, the prior distribution becomes
\begin{align}
    \mathcal{T}(\delta \mbf x) &= -\dfrac{1}{2} \big(\mbf x(\delta \mbf x)-\hat{\mbf x}^-\big)^T\mbf P^{-1}\big(\mbf x(\delta \mbf x)-\hat{\mbf x}^-\big) \\
    &=  -\dfrac{1}{2} \delta \mbf x^T \mbf P^{-1} \delta \mbf x
\end{align}

The measurement equation is represented as a polynomial of high order, with derivatives up to order $c$, using the Taylor expansion series by working directly onto the state polynomial:
\begin{equation}
    \mbf y (\delta \mbf x) = \mbf h (\mbf x(\delta \mbf x))
\end{equation}
This representation overcomes the limitation form previous flows that represent the measurement dependence in the flow equations merely via the Jacobian $\mbf H$, because every derivative is available thanks to the polynomial representation. Given the measurement outcome $\mbf y$ and the measurement noise covariance matrix $\mbf R$, the polynomial representation of the log likelihood can be evaluated as a function of the initial state deviation variable:
\begin{equation}
    \mathcal{L}(\delta \mbf x) = -\dfrac{1}{2} \big(\mbf y (\delta \mbf x)-\mbf y \big) ^T\mbf R^{-1}\big(\mbf y (\delta \mbf x)-\mbf y \big) 
\end{equation}

The particle flow can start by including the pseudo-time $\lambda$ to evaluate the posterior:
\begin{equation}
    \mathcal{P}(\delta \mbf x, \lambda) = \mathcal{T}(\delta \mbf x) + \lambda \mathcal{L}(\delta \mbf x) \label{post_DA}
\end{equation}
This is a mere summation of polynomials in the same variable, the particle deviation from their mean. Thus, the drift Eq. \eqref{drift_DA} and the diffusion Eq. \eqref{diff_DA} are applied directly onto the posterior polynomial, making differentiation easy and computationally fast. Moreover, according to the selected order of the Taylor expansion of the polynomial $\mbf y (\delta \mbf x)$, the differentiation goes beyond linearization. The gradient and Hessian of the posterior are evaluated 
\begin{align}
    \mathcal{G}(\delta \mbf x, \lambda) &= \nabla_{\delta \mbf x}\mathcal{P}(\delta \mbf x, \lambda)\\
    \mathcal{H}(\delta \mbf x, \lambda)&= \text{Hess}\big(\mathcal{P}(\delta \mbf x, \lambda)\big) = \nabla_{\delta \mbf x}\mathcal{G}^T(\delta \mbf x, \lambda)
\end{align}
as well as the gradient of the likelihood
\begin{equation}
    \mathcal{G}_{\mathcal{L}}(\delta \mbf x) =   \nabla_{\delta \mbf x}\mathcal{L}(\delta \mbf x)
\end{equation}
Each entry of gradient vectors and Hessian matrix is a polynomial in $\delta \mbf x$, whose value changes according to the particle's deviation. The drift and diffusion term can therefore be represented as a polynomial composition as well:
\begin{align}
    \mathcal{F}(\delta \mbf x, \lambda) &= \mathcal{H}(\delta \mbf x, \lambda)^{-1}\mathcal{G}_{\mathcal{L}}(\delta \mbf x) \label{drift_DA_fin}\\
    \mathcal{Q}(\delta \mbf x, \lambda) &= \mathcal{H}(\delta \mbf x, \lambda)^{-1}\mathcal{G}_{\mathcal{F}}(\delta \mbf x, \lambda) \label{diff_DA_fin}
\end{align}
where 
\begin{equation}
    \mathcal{G}_{\mathcal{F}}(\delta \mbf x) = \nabla_{\delta \mbf x}\mathcal{F}^T(\delta \mbf x, \lambda) 
\end{equation}
is the gradient of the drift polynomial vector. 

Equations \eqref{drift_DA_fin} and \eqref{diff_DA_fin} are the final drift and diffusion polynomials of the flow that represent the movement of each particle as a function of its distance from the mean (deviation) for each iteration of the pseudo-time $\lambda$ which represents the progress of the flow. To calculate numerically the drift vector and the diffusion matrix for each particle, the polynomials are evaluated at the given particle deviation, performing a simple variable numerical substitution: 
\begin{align}
    \mbf f(\mbf x^{(i)}, \lambda) &= \mathcal{F}(\delta \mbf x^{(i)}, \lambda) \label{drift_da_fin_v1}\\
    \mbf Q (\mbf x^{(i)},\lambda) &= \mathcal{Q}(\delta \mbf x^{(i)}, \lambda) \label{diff_da_fin_v1}
\end{align}
These two formulations are a more accurate representation of the Gromov flow reported in Eq. \eqref{drift_gro} and Eq. \eqref{diff_gro}, since the polynomials (centered at the prior mean) stop the expansion at a given order $c$ rather than with mere linearization. By increasing the expansion order indefinitely, the flow becomes more accurate since the polynomials get additional terms. Moreover, this approach saves an elevated number of operations that usually slow down common particle filters, since numerical integration is avoided. Indeed, the drift and diffusion polynomials need to be evaluated only once each $\lambda$ so that each particle's flow is calculated faster as polynomial evaluation rather than multiple floating point operations. 

For recursive estimation, a new prior mean and covariance are evaluated directly from the updated particles. When subject to a dynamical system, each particle can be propagated efficiently in the DA framework as in \cite{servadio2021differential,valli2012gaussian,servadio2024likelihood}.  

%%%%%%%%%%%%%%%%%%%%%%%%%%%%%%%%%%%%%%%%%%%%%%%%%%%%%%%%%%%%%%%%%%%%%%%%%%%%%%%%%%%%%%%%%
\subsection{The Particle as the Expansion Center - DAPFFv2-$c$}
The DA advantage to particle flow is the inclusion of high-order derivatives. However, the Taylor expansion is a local approximation of the measurement equation, meaning that as deviations become larger, the accuracy of the polynomials decreases. Therefore, a second approach can be studied to derive a particle flow in the DA framework, which is selecting each particle as the center of the Taylor expansion series. This selection evaluates polynomials for each particle, improving the accuracy of the evaluation of the derivatives. 

In this version of the DAPFF, one state polynomial per particle is created, centered at the particle location
\begin{equation}
    \mbf x^{(i)}(\delta \mbf x) =  \mbf x^{(i)} + \delta \mbf x
\end{equation}
The Gaussian prior still has its center at the particle mean, creating a polynomial per particle
\begin{align}
    \mathcal{T}^{(i)}(\delta \mbf x) &= -\dfrac{1}{2} \big(\mbf x^{(i)}(\delta \mbf x)-\hat{\mbf x}^-\big)^T\mbf P^{-1}\big(\mbf x^{(i)}(\delta \mbf x)-\hat{\mbf x}^-\big) 
\end{align}
This formulation is exact at the particle location rather than being exact at the mean as in DAPFFv1-$c$. In addition, the measurement expansion series up to order $c$ has its derivatives calculated at the particle location
\begin{equation}
    \mbf y^{(i)} (\delta \mbf x) = \mbf h (\mbf x^{(i)}(\delta \mbf x))
\end{equation}
creating a measurement polynomial per particle. Given the measurement and the noise covariance, the likelihood and the posterior follow the same derivation as the other version of the DA flow:
\begin{align}
    \mathcal{L}^{(i)}(\delta \mbf x) &= -\dfrac{1}{2} \big(\mbf y^{(i)} (\delta \mbf x)-\mbf y \big) ^T\mbf R^{-1}\big(\mbf y^{(i)} (\delta \mbf x)-\mbf y \big) \\
    \mathcal{P}^{(i)}(\delta \mbf x, \lambda) &= \mathcal{T}^{(i)}(\delta \mbf x) + \lambda \mathcal{L}^{(i)}(\delta \mbf x) \label{post_DA_v2}
\end{align}
Contrary to Eq. \eqref{post_DA}, there are as many expansions Eq. \eqref{post_DA_v2} as particles since, each time, the measurement polynomial is centered at the particle rather than at the mean.      

Given these representations of the distributions, the derivation of the drift and diffusion polynomials follows the same procedure as in the first version of the algorithm, but (once again) repeated each $i$-th particle:
\begin{align}
     \mathcal{G}^{(i)}(\delta \mbf x, \lambda) &= \nabla_{\delta \mbf x}\mathcal{P}^{(i)}(\delta \mbf x, \lambda)\\
    \mathcal{H}^{(i)}(\delta \mbf x, \lambda)&= \text{Hess}\big(\mathcal{P}^{(i)}(\delta \mbf x, \lambda)\big) \\
    \mathcal{G}^{(i)}_{\mathcal{L}}(\delta \mbf x) &=   \nabla_{\delta \mbf x}\mathcal{L}^{(i)}(\delta \mbf x) \\
    \mathcal{G}^{(i)}_{\mathcal{F}}(\delta \mbf x) &= \nabla_{\delta \mbf x}\mathcal{F}^{(i)T}(\delta \mbf x, \lambda)  
\end{align}
so that each particle has its own drift and diffusion polynomial
\begin{align}
    \mathcal{F}^{(i)}(\delta \mbf x, \lambda) &= \mathcal{H}^{(i)}(\delta \mbf x, \lambda)^{-1}\mathcal{G}^{(i)}_{\mathcal{L}}(\delta \mbf x) \label{drift_DA_fin_v2}\\
    \mathcal{Q}^{(i)}(\delta \mbf x, \lambda) &= \mathcal{H}^{(i)}(\delta \mbf x, \lambda)^{-1}\mathcal{G}^{(i)}_{\mathcal{F}}(\delta \mbf x, \lambda) \label{diff_DA_fin_v2}
\end{align}
In this case, to calculate the numerical drift vector and diffusion matrix for each particle, the constant term of the polynomial is extracted, which is equivalent to evaluating the polynomials at a null deviation vector:
\begin{align}
    \mbf f(\mbf x^{(i)}, \lambda) &=  \text{cons}(\mathcal{F}^{(i)}(\delta \mbf x, \lambda)) = \mathcal{F}^{(i)}(\mbf 0, \lambda) \label{drift_da_fin_v2}\\
    \mbf Q (\mbf x^{(i)},\lambda) &=  \text{cons}(\mathcal{Q}^{(i)}(\delta \mbf x, \lambda)) = \mathcal{Q}^{(i)}(\mbf 0, \lambda) \label{diff_da_fin_v2}
\end{align}
The DAPFFv2-$c$ is complete, and a new flow is implemented when a novel measurement becomes available. Similarly to DAPFFv1-$c$, if a propagation under a dynamical system is requested, each particle can be propagated rapidly using DA techniques until a new update is performed. 

This implementation if the DAPFF reduces to the Gromov flow when selecting $c=1$, meaning that DAPFFv2-$c$ can be considered as the high-order version of the Gromov flow, where the linearization assumption has been lifted in place of high-order derivatives. However, the benefits of calculating derivatives at the more accurate particle location comes at the cost of having a maximum order for this technique: $c=3$. In Eq. \eqref{drift_da_fin_v1} and \eqref{diff_da_fin_v1}, the expansion can go up to any arbitrary order, as it is evaluated at a specific deviation: the center at the mean allows an arbitrarily high order of the expansions. In Eq. \eqref{drift_da_fin_v2} and \eqref{diff_da_fin_v2}, the coefficients of the expansions are more accurate since they are locally evaluated centered at the particle, but it restricts the maximum improvement in order up to 3 since the polynomials are evaluated at a null deviation. Any higher order would compute bringing no benefit, as the additional monomials in the expansion would be neglected being multiplied by zero.  

%%%%%%%%%%%%%%%%%%%%%%%%%%%%%%%%%%%%%%%%%%%%%%%%%%%%%%%%%%%%%%%%%%%%%%%%%%%%%%%%%%%%%%%%%
\subsection{Integration of the Flow}
The flow can be computed following the integration technique that best fits the application, as the drift and diffusion parameters of the stochastic differential equation are provided. The common selection of the explicit strong Euler-Maruyama method \cite{platen2010numerical} is here derived, and it has been selected for the numerical applications reported in this paper. 

The Euler-Maruyama method uses an integration step of $\Delta \lambda$ to move from pseudo-time $\lambda$ to $\lambda + \Delta \lambda$, integrating, recursively, each particle according to 
\begin{equation}
    \mbf x^{(i)}_{\lambda + \Delta \lambda} = \mbf x^{(i)}_{\lambda}\mbf f\big(\mbf x^{(i)}_{\lambda}, \lambda\big) + \mbf B \big(\mbf x^{(i)}_{\lambda},\lambda \big) \tilde{\mbf w}
\end{equation}
where $\mbf B (\mbf x^{(i)}_{\lambda},\lambda)$ is calculated via the Cholesky decomposition of $\mbf Q (\mbf x^{(i)}_\lambda,\lambda)$ and $\tilde{\mbf w}$ is a zero-mean Gaussian random variable with covariance matrix $\Delta\lambda \mbf I$.

When considering diffusion, the decomposition of $\mbf Q (\mbf x^{(i)}_\lambda,\lambda)$ is the mathematically most challenging step, regardless of the methodology selected to obtain the diffusion term. In fact, matrix $\mbf Q (\mbf x^{(i)}_\lambda,\lambda)$ is usually singular and an accurate triangular decomposition becomes unavailable. An LDL decomposition can be implemented to aid this issue, where  $\mbf Q = \mbf L \mbf T \mbf L^T$ so that $\mbf B = \mbf L \mbf T^{1/2}$, where $\mbf T$ is a diagonal matrix such that the root operator can be applied element-wise.

%%%%%%%%%%%%%%%%%%%%%%%%%%%%%%%%%%%%%%%%%%%%%%%%%%%%%%%%%%%%%%%%%%%%%%%%%%%%%%%%%%%%%%%%%
%%%%%%%%%%%%%%%%%%%%%%%%%%%%%%%%%%%%%%%%%%%%%%%%%%%%%%%%%%%%%%%%%%%%%%%%%%%%%%%%%%%%%%%%%
%%%%%%%%%%%%%%%%%%%%%%%%%%%%%%%%%%%%%%%%%%%%%%%%%%%%%%%%%%%%%%%%%%%%%%%%%%%%%%%%%%%%%%%%%
\section{Numerical Application} 
The two versions of the DAPFF have been tested on a numerical example presented in \cite{michaelson2024particle} to highlight the benefits and drawbacks of particle flow filters. Consider a prior state estimate 
\begin{equation}
    \hat{\mbf x}^{-} = \begin{bmatrix}
        -3.5 & 0 
    \end{bmatrix}^T
\end{equation}
with error covariance 
\begin{equation}
    \mbf P^{-} = 
    \begin{bmatrix}
        1 & 0.5 \\ 0.5 & 1
    \end{bmatrix}
\end{equation}
A range measurement is given as
\begin{equation}
    y = || \mbf x|| + v
\end{equation}
where $v$ is the measurement noise, with distribution $v \sim \mathcal{N}(0,0.1^2)$. The numerical outcome received from the filter is $y = 1$, and it is desired to update the distribution, obtaining a representation of the true posterior using a particle flow approach. 

\begin{figure}[!htb]
    \centering
    \includegraphics[width=1.0\linewidth]{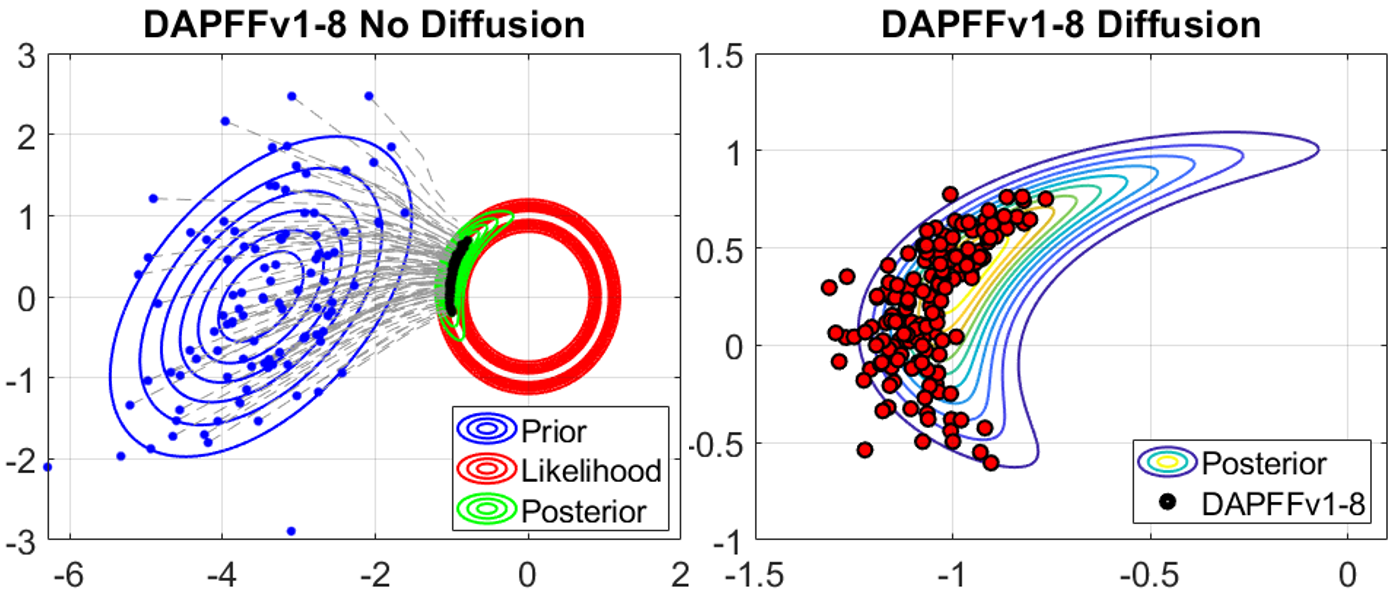}
    \caption{DAPFFv1-8 Solution with Diffusion.}
    \label{fig:dapffv1}
\end{figure}

Given the range measurement, the likelihood distribution is of circular space, with its maximum at the unity circle and probability decreasing getting radially farther from it. Given a Gaussian prior far from the measurement, the resulting posterior highlights the region of the likelihood closer to the prior. Figure \ref{fig:dapffv1} (left) shows the three distributions in blue (prior), likelihood (red), and green (posterior). The particle flow filters are applied to this scenario after sampling particles directly from the prior, meaning that each particle holds the same weight. During the flow integration according to the Euler-Maruyama method, a pseudo-time step of $\Delta\lambda=1/50$ has been applied. 

Figure \ref{fig:dapffv1} shows the particle flow of the DAPFFv1-8 solution. As a reminder, the number 8 in the name of the estimator means that polynomials have monomials up to 8th order in the expansion. The left plot in the figure reports the particles in their initial state sampled from the prior in blue, $\lambda = 0$, while the black particles show their position at the end of the flow integration, $\lambda = 1$. The gray dashed lines highlight the pathways followed by the particles and how they migrate toward the region of the high probability of the posterior distribution. The diffusion term has been turned off for this representation in order to show the drift effect and analyze the particle movement. When diffusion is considered, the particle randomly jumps from the marked pathways due to the stochastic influence of $\tilde{\mbf w}$. Therefore, Fig. \ref{fig:dapffv1} (right) reports the particles already in their final position when diffusion is included in the integration. The diffusion term helps the particles spread around the region of highest likelihood, trying to cover the full posterior distribution. It can be noted that the particles tend to lay on the outer rim of the posterior, neglecting the inner one. This is due to the numerical instability of the $\mbf Q$ decomposition, which gives numerical issues when a particle crosses the unity circle. 

\begin{figure}[!htb]
    \centering
    \includegraphics[width=1.0\linewidth]{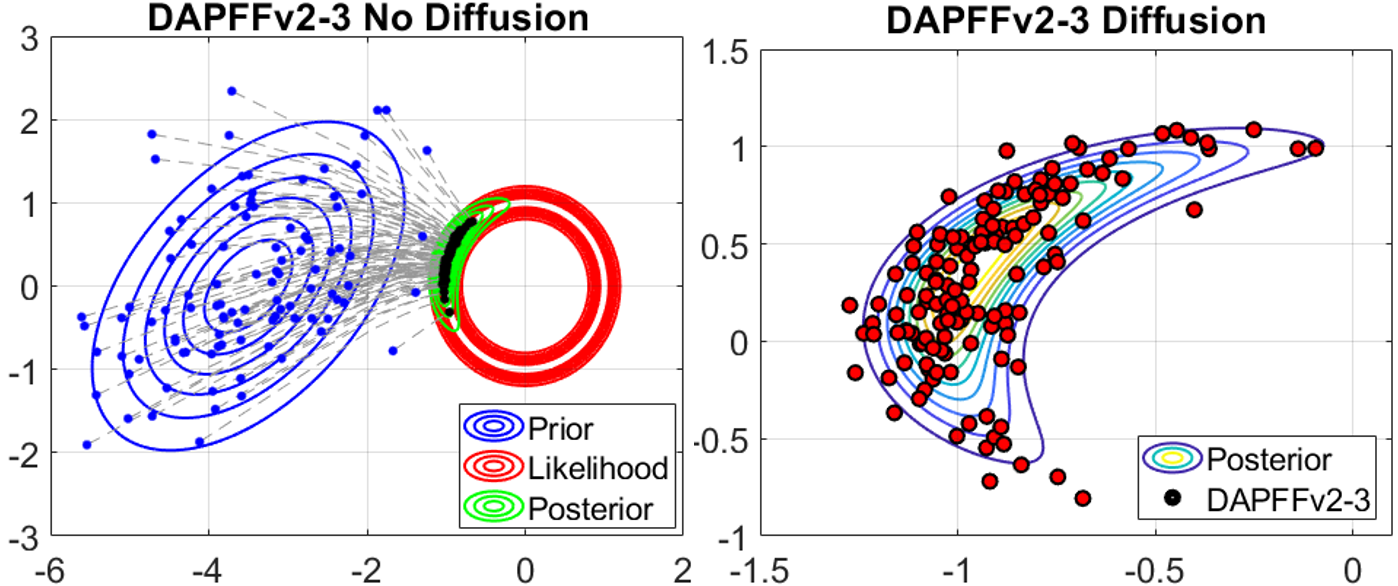}
    \caption{DAPFFv2-3 Solution with Diffusion.}
    \label{fig:dapffv2}
\end{figure}
The same application has been solved using the DAPFFv2-3 and reported in Figure \ref{fig:dapffv2}. Once again, the number 3 in the name of the estimator means that a third-order expansion has been performed. From the left plot of the figure, it can be seen how the particles correctly migrate toward the posterior, following a smooth and continuous trajectory due to the lack of diffusion. When diffusion is included (right side of the figure) the particles spread accordingly to the width of the posterior, covering it accurately according to the regions of low and high probability density. 

The DAPFF algorithm, in both its versions, provides accurate results even when the acquired measurement is far from the region highlighted by the prior, as shown in the example, an issue that would break common particle filters such as the bootstrap particle filter. The DAPFF has therefore being compared to the exact flow and to the Gromov flow. The best way to analyze the performance of the filters is to study their drift component, as it dictates the overall push of the flow, while diffusion is a deviation from it to account for the spread of the final posterior. 

\begin{figure}[!htb]
    \centering
    \includegraphics[width=1.0\linewidth]{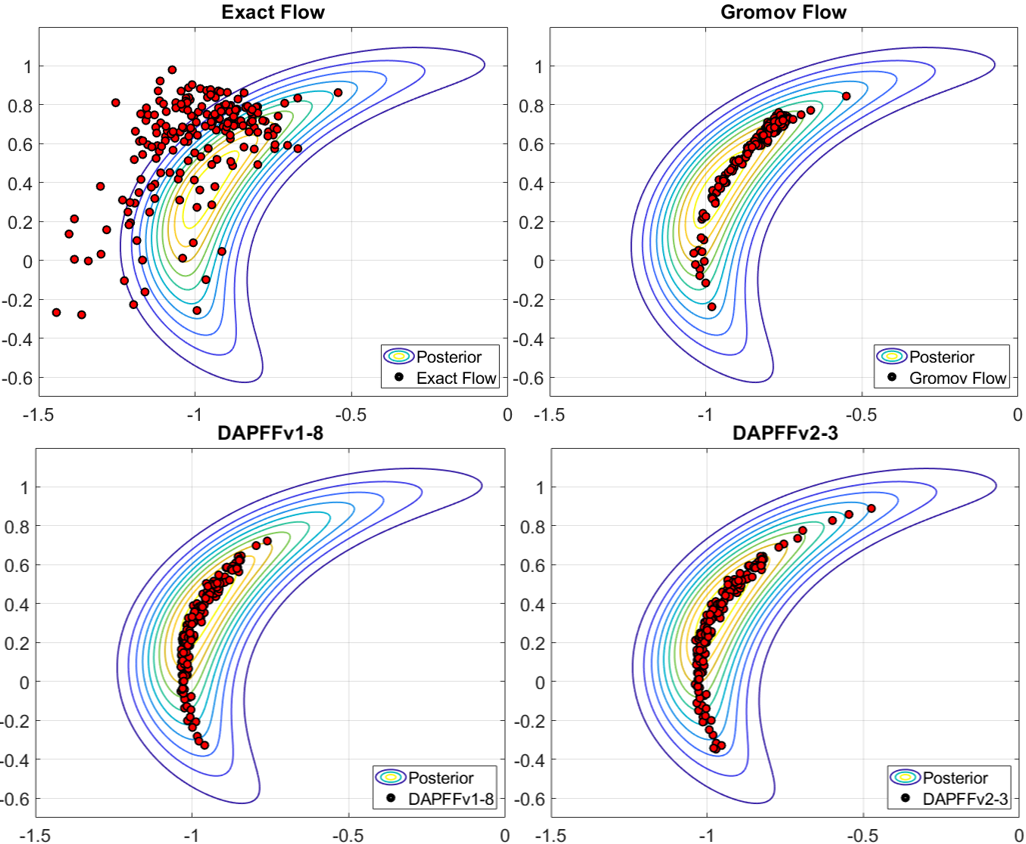}
    \caption{Posterior representation with four separate Particle Flow Filters without Diffusion}
    \label{fig:final}
\end{figure}
Figure \ref{fig:final} shows the final ensemble of particles (200 particles) evaluated using the four different kinds of flow, evaluated using only the drift term. For fairness in the comparison, the same initial random set of particles drawn for the prior Gaussian has been used for each methodology. The exact flow solution is the least accurate, as the particles agglomerate around the posterior in an unorganized manner, not representing the PDF properly. On the contrary, the Gromov flow is able to capture part of the curvature of the distribution. However, most of the particles are condensed in a small space, and only a few align correctly on the peak of the distribution. Moreover, the location of this agglomerate is not the region with the maximum density, creating an inaccurate estimate. The plots in the bottom row report the DAPFF, version 1 on the left and version 2 on the right. They both show a consistent and accurate behavior of the drift of the flow, placing particles on the crest of the distribution, following the curve of the PDF. This is the best outcome, as more particles are in the high density region, and few get farther away to describe the spread, following the likelihood correctly. The DAPFF has, therefore, the more accurate drift term among the selected filters since, once the diffusion term is included, the stochasticity of the Wiener process will account for the spread of the particle from the crest. The high-order polynomials add information about the shape of the likelihood function and its curvature, enabling the flow to address the nonlinearity of the posterior distribution better. Indeed, the increase of the expansion order $c$ correlates to a higher filter accuracy, as shown in the following analysis.

\begin{figure*}[!htbp]
    \centering
    \includegraphics[width=1.0\linewidth]{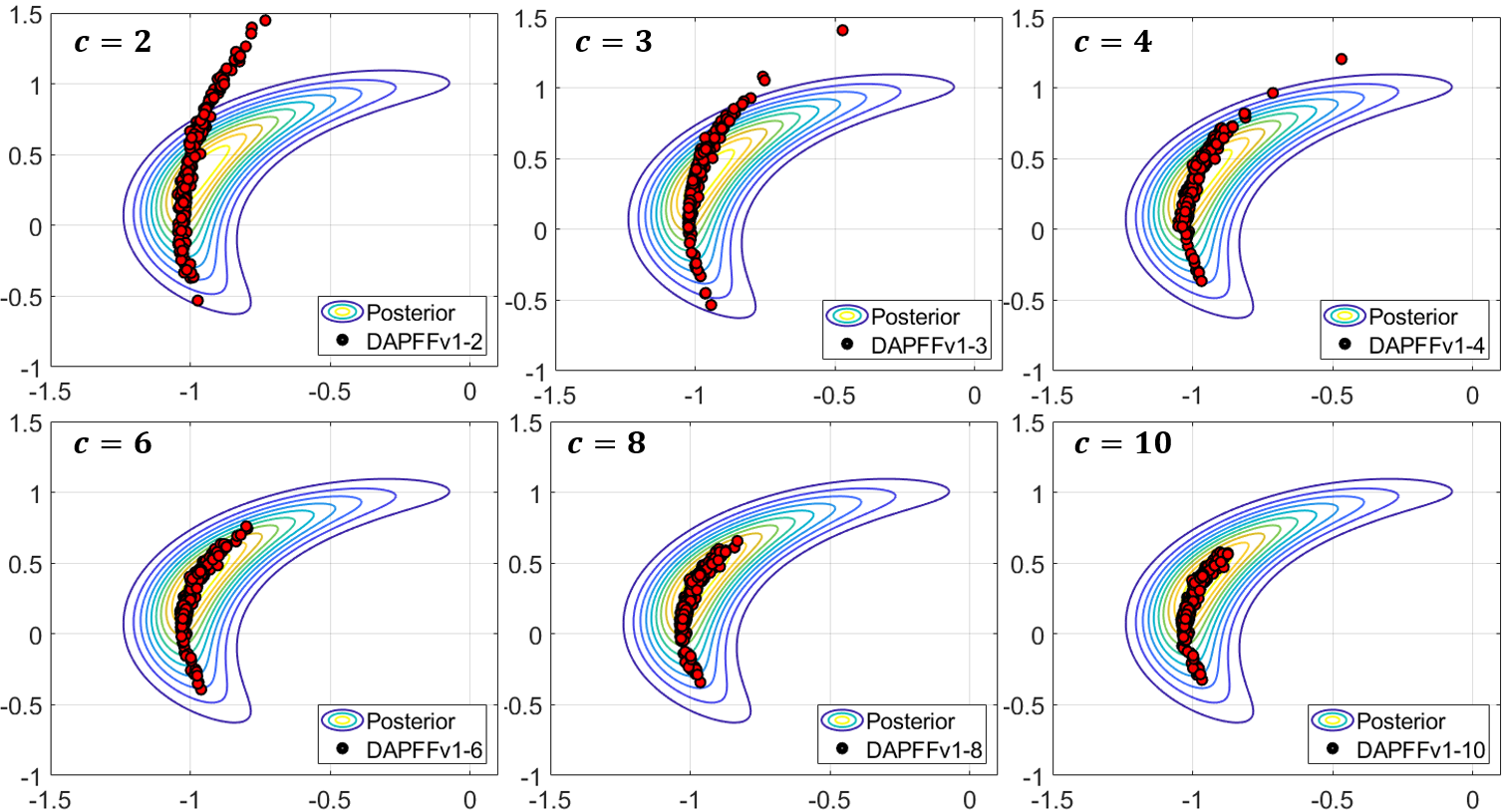}
    \caption{DAPFFv1-$c$ Expansion Order Analysis with $c =  [2 \ \ 3 \ \ 4 \ \ 6 \ \ 8 \ \ 10]$.}
    \label{fig:dapffv1_an}
\end{figure*}
The influence of the selection of the expansion order is reported in Fig. \ref{fig:dapffv1_an}, where the range estimation problem has been repeated with DAPFFv1-$c$ for orders spanning 2 to 10. DAPFFv1-$c$ has the expansion centered at the mean so that the flow of each particle is evaluated depending on the deviation of each particle from the mean itself. This feature allows an arbitrarily high order of the series, as the polynomials are evaluated with a not-null deviation. Looking at the figure, increasing the order better approximates the curvature of the distribution. Once again, for representation purposes, the diffusion term has been neglected to highlight the benefits the new techniques bring to the flow derivation. The expansion order increase adds information to the polynomials, meaning that they better account for the system's nonlinearities. Indeed, moving from order 2 up to 10, the particles show a more accurate distribution and an approximation of the posterior. Order 2 in DAPFFv1-$c$ is the minimum allowed, as there are two differentiations in the evaluation of the flow, meaning that initial monomials of the second order are required to have a non-zero contribution.

\begin{figure*}[!htbp]
    \centering
    \includegraphics[width=1.0\linewidth]{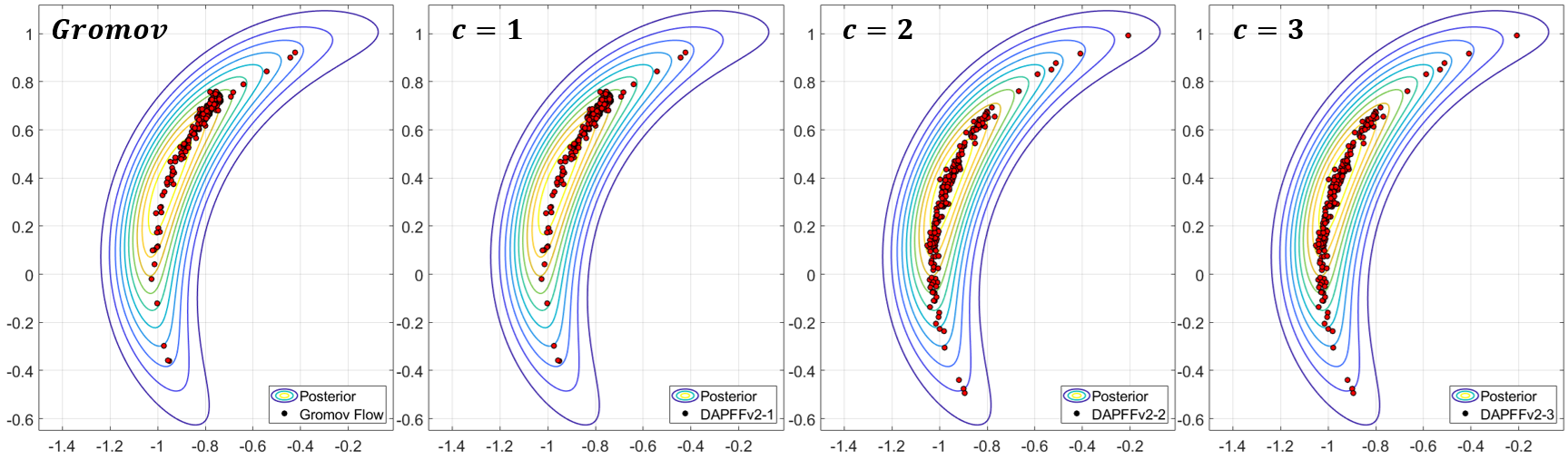}
    \caption{DAPFFv2-$c$ Expansion Order Analysis with $c =  [1 \ \ 2 \ \ 3]$.}
    \label{fig:dapffv2_an}
\end{figure*}
A similar analysis has been performed with DAPFFv2-$c$, where the expansion order has been changed along its full range, from 1 to 3. Since DAPFFv2-$c$ has the polynomials centered at each particle's location, increasing the order above 3 would bring no benefit to the algorithm, as the expansions are evaluated at null deviations, and $c>3$ would add null contributions to the evaluation. This version of the DAPFF is the closest to classic particle flow filters, as the flow is optimized for the particles. Figure \ref{fig:dapffv2_an} shows improvement and similarities in the expansion order study. Starting each update with the same initial random draw of particles from the prior, it can be noted that DAPFFv2-1 reduces to the Gromov flow. The particles' final location is the same, and the two filters are identical. They both provide a faulty ensemble of particles since most of them stop away from the region of maximum density in the PDF. Increasing the order leads to a more accurate representation, as shown in the $c=2$ and $c=3$ cases. Here, the particles are better spread according to the posterior with respect to the $c=1$ case, showing no erroneous skewed behavior. However, there is no difference in upgrading the order from second to third in this drift-only case, as the third differentiation comes from the evaluation of the diffusion term in the gradient of the drift term. Indeed, if diffusion is neglected, the two orders yield the same result. The gain in accuracy is due to the capability of the expansion to include curvature in the approximation (conceptually, being able to follow a parabola rather than a line). Once a parabolic behavior is achieved, the flow can follow a more accurate representation of the function, showing significant benefits upgrading to parabolic from linearity, as expected. 
\begin{figure}[!htb]
    \centering
    \includegraphics[width=1.0\linewidth]{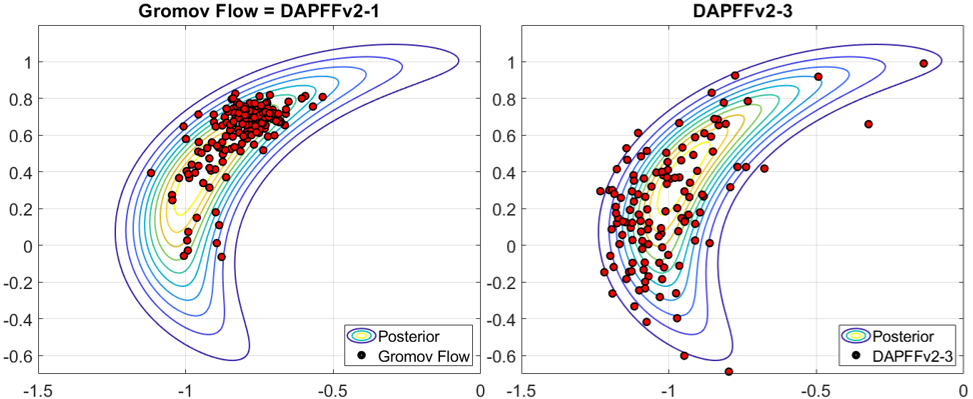}
    \caption{Gromov vs DAPFFv2-3 Solution.}
    \label{fig:gvsd}
\end{figure}
When diffusion is added back into the estimation, Fig. \ref{fig:gvsd} shows how the Gromov flow has all the particles merged together, while the DAPFFv2-3 better describes the true posterior distribution. In this presented application, a better representation of the posterior, thanks to an adequate spread of the particles, leads to a more reliable estimate. Indeed, while the mean of the particles in both filters might give a closer error range, the DAPFFv2-3 better predicts the error covariance, making the technique suitable for recursive estimation in filtering, where each particle will undergo propagation.

%%%%%%%%%%%%%%%%%%%%%%%%%%%%%%%%%%%%%%%%%%%%%%%%%%%%%%%%%%%%%%%%%%%%%%%%%%%%%%%%%%%%%%%%%
%%%%%%%%%%%%%%%%%%%%%%%%%%%%%%%%%%%%%%%%%%%%%%%%%%%%%%%%%%%%%%%%%%%%%%%%%%%%%%%%%%%%%%%%%
%%%%%%%%%%%%%%%%%%%%%%%%%%%%%%%%%%%%%%%%%%%%%%%%%%%%%%%%%%%%%%%%%%%%%%%%%%%%%%%%%%%%%%%%%
\section{Conclusion}
The classic derivation of the particle flow filter has been revised and improved in the DA framework considering high-order terms in the Taylor expansion. Thanks to a detailed representation of the measurement equation in the derivation of the drift and diffusion terms, the particle travels more accurately in the flow, representing the true posterior distribution better than Gromov's flow or the exact flow. The derivation of the flow terms follows the main definition of particle flow, but it is applied directly to the polynomial representation of the distribution, such that differentiation can be performed effectively in the DA framework. 

In the novel derivation, the center of the polynomial expansion can be picked according to two separate options: either at the particle or at the mean of the prior distribution. This selection drives the two versions of the DAPFF and their different way of evaluating polynomials due to the different definitions of deviation vectors. The numerical application showed the improvement gained by the high-order monomials in the expansions and opens these techniques to complete filtering techniques with embedded prediction and PDF propagation.

%%%%%%%%%%%%%%%%%%%%%%%%%%%%%%%%%%%%%%%%%%%%%%%%%%%%%%%%%%%%%%%%%%%%%%%%%%%%%%%%%%%%%%%%%
%%%%%%%%%%%%%%%%%%%%%%%%%%%%%%%%%%%%%%%%%%%%%%%%%%%%%%%%%%%%%%%%%%%%%%%%%%%%%%%%%%%%%%%%%
%%%%%%%%%%%%%%%%%%%%%%%%%%%%%%%%%%%%%%%%%%%%%%%%%%%%%%%%%%%%%%%%%%%%%%%%%%%%%%%%%%%%%%%%%
\bibliographystyle{ieeetr}
\bibliography{references.bib}

\end{document}